\documentclass[english,English]{article}
\usepackage[latin9]{inputenc}
\usepackage{amsmath}
\usepackage{graphicx}

\makeatletter

\providecommand{\tabularnewline}{\\}

\usepackage[section]{placeins}

\providecommand{\tabularnewline}{\\}

\usepackage{tikz}
\usepackage{verbatim}
\usetikzlibrary{arrows}
\usepackage{caption}
\usepackage[english]{babel}
\usepackage{floatrow}
\providecommand{\tabularnewline}{\\}

\providecommand{\tabularnewline}{\\}

\usepackage{babel}

\usepackage{babel}

\makeatother

\usepackage{babel}
\begin{document}

\title{The Q.Q Interaction and Variations of Single Particle Energies}

\author{Mingyang Ma$^{1}$, Praveen Srivastava$^{2}$ and Larry Zamick$^{1}$
\\
 1. Department of Physics and Astronomy, \\
 Rutgers University, Piscataway, New Jersey 08854\\
 2. Department of Physics,\\
 Indian Institute of Technology Roorkee, Roorkee 247 667, India}

\maketitle

\section{Abstract}

In this work we make further studies of the quadruple quadruple interaction
used in shell model calculations.Whereas in a previous work we adjusted
the single particle energies so as to obtain the rotational spectrum
of the Elliott model, we here vary the single particle energies and
examine the various spectral shapes that evolve.

\section{Introduction}

In a previous work {[}1{]} we used a Q.Q interaction to study the
spectrum of $^{20}$Ne in the context of a shell model calculation.
To get the much studied Elliott model results {[}2-7{]}, namely of
a $J(J+1)$ spectrum , one has to choose the single particle energies
with care.\\
 The Elliott formula {[}2{]} for the energies is 
\begin{equation}
E(SU(3))=\chi'[-4(\lambda^{2}+\mu^{2}+\lambda\mu+3(\lambda+\mu))]+3\chi'L(L+1)
\end{equation}
where $\chi$'= 5b$^{4}$/(32$\pi$) $\chi$.\\
 To get these SU(3) results in the shell model one has to introduce
a single particle energy splitting {[}8, 9{]} 
\begin{equation}
E(L_{2})-E(L_{1})=3\chi'[L_{2}(L_{2}+1)-L_{1}(L_{1}+1)]
\end{equation}
We can now use the simple Q.Q interaction without the momentum terms.
For completeness we also list the rotational model formulas of Bohr
and Mottelson {[}10{]}. 
\begin{equation}
B(E_{2}(I_{i}K\rightarrow I_{f}K)=(5/16\pi)Q_{0}^{2}(I_{i}K2\;0|I_{f}K)^{2}.
\end{equation}
\begin{equation}
Q(I,K)=(3K^{2}-I(I+1))/((I+1)(2I+3))Q_{0}.
\end{equation}
It should be noted that although both the Elliott model and the rotational
model yield $J(J+1)$ spectra, they do not agree on the $B(E2)$'s.
For a $J$ to $(J-2)$ transition, the Elliott model shows a decreased
$B(E2)$ with increasing $J$ and band termination whereas with Bohr
and Mottelson, the corresponding $B(E2)$s increase with increasing
$J$.\\
 What is new in this work is that we will change the single particle
energies.We will then of course no longer get rotational spectra but
it will be of interest to see what new spectral shapes emerge.

\section{The 0p shell.}

\noindent If we make the 0p$_{3/2}$ and 0p$_{1/2}$ single particle
energies degenerate ,then with the Q.Q interaction we get ``rotational
band'' energies for the yrast $J=0,2,4$ states --- $E(J)=CJ(J+1)$.
In Table 1 we show the $p$ shell yrast spectrum as we introduce a
single particle splitting $e(p_{1/2})-e(p_{3/2})=\Delta$. The results
are also shown in Fig 1. 
\begin{table}[h]
\centering %
\begin{tabular}{|c|c|c|c|c|c|}
\hline 
$J/\Delta$  & 0  & 1  & 5  & 10  & 100\tabularnewline
\hline 
\hline 
0  & 0  & 0  & 0  & 0  & 0\tabularnewline
\hline 
2  & 0.8953  & 0.9122  & 0.8753  & 0.8416  & 0.7965\tabularnewline
\hline 
4  & 2.9844  & 2.8967  & 2.2777  & 2.1340  & 1.9912\tabularnewline
\hline 
$E(4)/E(2)$  & 3.3333  & 3.1777  & 2.5930  & 2.5362  & 2.5\tabularnewline
\hline 
\end{tabular}\caption{Energy Levels for the Configuration $(p^{2})_{\pi}$ and $(p^{2})_{\mu}$
as a Function of $\Delta=e(p_{1/2})-e(p_{3/2})$. The Q.Q Interaction
Is Used}
\label{tab:my_label} 
\end{table}

\begin{figure}[h]
\centering \includegraphics[width=11.2cm]{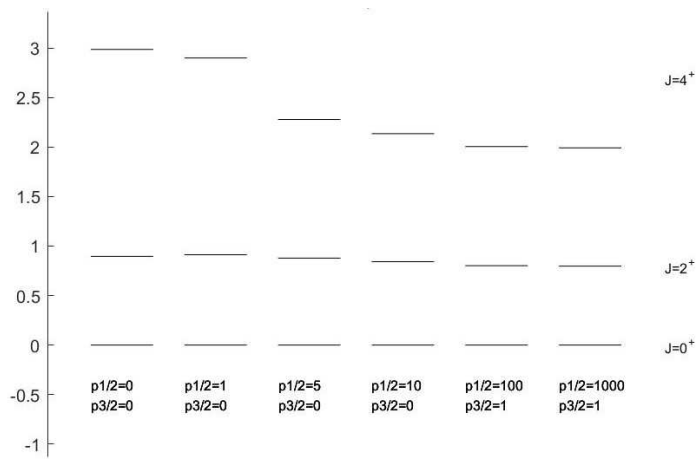}
\caption{P Shell Yrast Spectrum }
\label{fig:my_label} 
\end{figure}

\FloatBarrier Note that the ratio $E4/E2$ starts at 3.3333 and ends
at 2.5. The beginning is of course the rotational limit, as was shown
by Elliott. The ratio 2.5 corresponds to the single $j$ shell case
--- pure $p_{3/2}$. In this limit the quadruple moment of the 2$^{+}$state
is zero because we are at midshell.\\
 Amusingly, in a recent publication by Sharon et al.{[}11{]} a similar
journey was described from a strong prolate deformation to a gamma
soft rotor{[}12{]}. At the latter $O(6)$ limit {[}13{]}, the quadruple
moment was also zero and the ratio of energies $E4/E2$ was equal
to 2.5.

\noindent The expression for the matrix elements for 2 protons and
2 neutrons in a single j shell is here given.

$<$ (j j)$^{J_{p}}$ (j j)$^{J_{n}}$ $|V|$ (jj)$^{J_{p}}$$^{'}$
(j j)$^{J_{n}}$$^{'}$$>$ $^{J}$=

4($\sum$((j j)$^{J_{p}}$ (j j)$^{J_{n}}$ $|$ ( j j)$^{J_{a}}$
(j j)$^{J_{b}}$))$^{J}$ {*}((j j)$^{J_{p}}$$^{'}$(j j)$^{J_{n}}$$^{'}$
$|$( j j)$^{J_{a}}$ (j j)$^{J_{b}}$))$^{J}$ E(J$_{b}$) + (E(J$_{p}$)
+ (E(J$_{n}$)) $\delta$$_{J_{p}}$$_{J_{p}}$$'$ $\delta$$_{J_{n}}$$_{J}$$_{n}$'.

In the above we use unitary 9j symbols-schematically

U9j= ($(2J_{p}+1)(2(J_{n}$ +1) (2J$_{a}$+1) (2J$_{b}+1)$)$^{1/2}$
regular 9j. Here Jp, Jn, Jp', Jn' and Jb are all even. Ja can be even
or odd.

We next look at the 2 body matrix elements V(J) of Q.Q in the p$_{3/2}$
shell. From J=0 to J=3 they are \{-0.497359, -0.099472,-0.298416 and
-0.099472\}

$\chi$ b$^{4}$.

At first these look very complicated but note that E(1) equals E(3).
Let us add a constant to all of these matrix elements so that E(1)=E(3)=0.
This will not affect the spectrum. Thus the shifted levels are \{
-EY, 0, $EY,0$\} $\chi$ b$^{4}$ with $EY$ = 0.39789.

We now address the problem of why the ratio E(4)/E(2) =2.5. The basics
states are {[}J$_{p}$ J$_{n}${]}J.

There is only one configuration for J=4 {[}2 2{]}4. Using units of
EY $\chi$ b$^{4}$ we find E(4)= 3.

For J=0 we have 2 configurations {[}0 0{]}0 and {[}2 2{]}0. The secular
matrix is 

\begin{center}
\begin{tabular}{|c|c|}
\hline 
-1  & -2.236068\tabularnewline
\hline 
\hline 
-2.236068  & 3\tabularnewline
\hline 
\end{tabular}
\par\end{center}

The eigenvalues are -2, and 4.

For J=2 we have 3 configurations {[}0 2{]}2 {[}2 0{]}2 and {[}2 2{]}2.
The secular matrix

is 

\begin{center}
\begin{tabular}{|c|c|c|}
\hline 
1  & 1  & 0\tabularnewline
\hline 
\hline 
1  & 1  & 0\tabularnewline
\hline 
0  & 0  & 3\tabularnewline
\hline 
\end{tabular}
\par\end{center}

The eigenvalues are 0, 2 and 3.

Thus the lowest eigenvalues for J=0,2 and 4 are respectively -2, 0
and 3. If we again make a shift so that the J=0 state is at zero energy
then we have

E'(0)= 0, E'(2)= 2 and E'(5)=5 and hence E'(5)/E'(2) =2.5.

Note that there is no coupling between {[}0 2 {]}2 and {[}2 2{]}2,
or between {[}2 0 {]}2 and {[}2 2{]}2. This is due the the vanishing
of the U9j

((3/2 3/2)0 ,(3/2 3/2)2 \textbar{} (3/2 3/2)2, (3/2 3/2) 2)$^{2}$.
The vanishings of certain 6j and 9j coefficients has been previously
studied by Robinson and Zamick {[}14{]} and references therein.

\section{The 1s 0d shell.}

\noindent As noted by Kingan et al. {[}1{]} if, with the Q.Q interaction,
we choose the single particle energies in the $1s-0d$ shell such
that $e(0d_{3/2})=e(0d_{5/2)})=0.8952$ and $e(1s_{1/2})=0$, we get
a rotational band. Starting from there we now move the $1s$ above
the degenerate $d$ pair by an amount $\Delta$ and explore what interesting
spaces evolve . We show numerical results in Table 2 and visual results
in Fig. 2. 
\begin{table}[h]
\centering %
\begin{tabular}{|c|c|c|c|c|c|c|c|}
\hline 
J/$\Delta$  & Rot  & 3  & 8  & 9  & 10  & 20  & 100\tabularnewline
\hline 
\hline 
0  & 0  & 0  & 0  & 0  & 0  & 0  & 0\tabularnewline
\hline 
2  & 0.8952  & 0.6469  & 1.1499  & 1.2420  & 1.3199  & 1.2075  & 0.7264\tabularnewline
\hline 
4  & 2.9840  & 2.0062  & 1.3372  & 1.2999  & 1.2701  & 1.1362  & 1.0247\tabularnewline
\hline 
6  & 6.2664  & 4.2063  & 3.2534  & 3.2486  & 3.2532  & 3.3453  & 2.9701\tabularnewline
\hline 
8  & 10.7424  & 6.8472  & 3.7202  & 3.5202  & 3.3587  & 2.6402  & 2.1115\tabularnewline
\hline 
$E(4)/E(2)$  & 3.3333  & 3.1013  & 1.1629  & 1.0466  & 0.9623  & 0.9410  & 1.4107\tabularnewline
\hline 
\end{tabular}\caption{Separating $1s$ From $0d$ by an Amount $\Delta$ for 2 Protons and
2 Neutrons in the $1s-0d$ Shell. The Q.Q Interaction Is Used.}
\label{tab:my_label} 
\end{table}

\begin{figure}[h]
\centering \includegraphics[width=12cm]{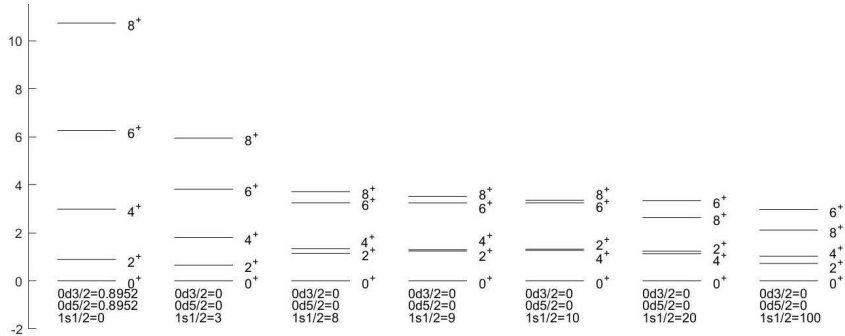}
\caption{Separating $1s$ From $0d$ by an Amount $\Delta$ for 2 Protons and
2 Neutrons in the $1s-0d$ Shell}
\label{fig:my_label} 
\end{figure}

As shown in figure 2, an interesting behavior emerges near $\Delta=9$.
One gets 2 sets of near doublets with the $J=2^{+}$ and $4^{+}$
nearly degenerate and likewise the $6^{+}$ and $8^{+}$. Note also
that from $\Delta$=10 to 20, the $2^{+}$state is at a higher energy
than the $4^{+}$state. However for very large $\Delta$, the usual
order recovers with $J=4^{+}$ higher than $J=2^{+}$. However the
$J=8^{+}$ state comes below the $J=6^{+}$ state causing $J=8^{+}$
to be isomeric. Such behavior is not unusual.\\

\begin{figure}[h]
\centering \includegraphics[width=12cm]{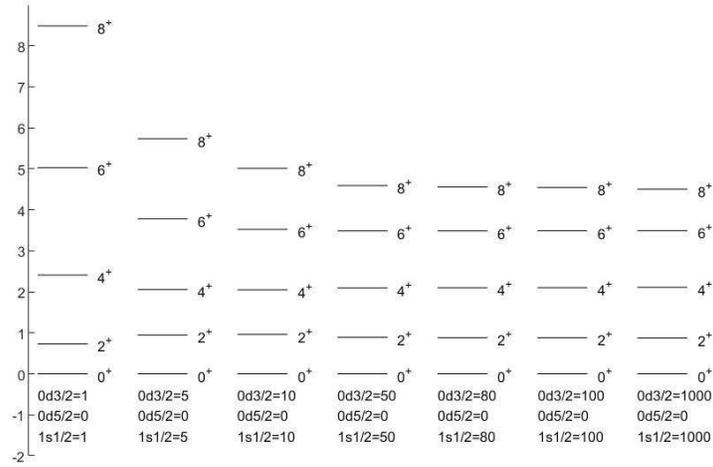}
\caption{The $1s_{1/2}$ and $0d_{3/2}$ Are Set Equal and Moved an Amount
$\Delta$ Above $0d_{5/2}$}
\label{fig:my_label} 
\end{figure}

\FloatBarrier The trend in Fig 3, where both the $1s_{1/2}$ and
$0d_{3/2}$ single particle energies are raised relative to $0d_{5/2}$
is simpler than in Fig.2. Note that the ratio $E4/E2$,which in the
rotational limit is $10/3$, has a value of 3.298 for $\Delta$=1
whereas when $\Delta$ is very large the value reduces to 2.417. The
latter is not too far off from the value of 2.5 in the p shell.

\section{Added Remarks}

We have had a long standing interest in results emanating from schematic
interactions and in particular those from the Q.Q interaction. For
example ,with Esduderos {[}15,16{]} interesting degeneracies were
found which involved isospin $T=0$ and $T=2$ ''doublets''. With
Harper {[}17{]} it was found that the wave functions of a single $j$
shell for a system of 2 protons and 2 neutrons could be well approximated
by unitary $9j$ coefficients.Generally speaking, schematic interactions
can give simple physical insights to more complex behaviors.\\
 The calculations in this work were carried out using The Shell Model
Code NUSHELLX@MSU{[}18{]}.

\end{document}